# Investigating the Proton Structure:
# The FAMU experiment

A. Vacchi[1,5,20], A. Adamczak[2], D. Bakalov[3], G. Baldazzi[4], M. Baruzzo[1,5], R. Benocci[6,7], R. Bertoni[6], M. Bonesini[6,8], H. Cabrera[1,17], S. Carsi [6,22], D. Cirrincione[1,5], F. Chignoli[6], M. Clemenza[6,8], L. Colace[9,10], M.Danailov[1,11], P. Danev[3], A. de Bari[13], C. De Vecchi[13], M. De Vincenzi[9,14], E. Fasci[15], K. S. Gadedjisso-Tossou [1,17,18], L. Gianfrani[15], A. D. Hillier[19], K. Ishida[19,20], P. J. C. King[19], V. Maggi[6,7], R. Mazza[6], A. Menegolli[12,13], E. Mocchiutti[1], L. Moretti [15,8], G. Morgante[4,16], J. Niemela[17], C. Petroselli[6,28], C. Pizzolotto[1], A. Pullia[22,24], R. Ramponi[23,25], H. E. Roman[8], M. Rossella[13], R. Rossini [12,13], R. Sarkar[26], A. Sbrizzi[4], M. Stoilov[3], L. Stoychev[3], J. J. Suarez-Vargas[1], G. Toci[21], L. Tortora[9], E. Vallazza[6], C. Xiao[28], K. Yokoyama[19]

1 Sezione INFN di Trieste, via A. Valerio 2, Trieste, Italy
2 Institute of Nuclear Physics, Polish Academy of Sciences, Radzikowskiego 152, PL31342 Kraków, Poland
3 Institute for Nuclear Research and Nuclear Energy, Bulgarian Academy of Sciences, blvd. Tsarigradsko ch. 72, Sofia 1784, Bulgaria
4 Sezione INFN di Bologna, viale Berti Pichat 6/2, Bologna, Italy
5 Università di Udine, via delle Scienze 206, Udine, Italy
6 Sezione INFN di Milano Bicocca, Piazza della Scienza 3, Milano, Italy
7 DISAT Università di Milano Bicocca, Piazza della Scienza 1, Milano, Italy
8 Dipartimento di Fisica G. Occhialini, Università di Milano Bicocca, Piazza della Scienza 3, Milano, Italy
9 Sezione INFN di Roma Tre, Via della Vasca Navale 84, Roma, Italy
10 Dipartimento di Ingegneria, Università degli Studi Roma Tre, Via V. Volterra 62, Roma, Italy
11 Sincrotrone Elettra Trieste, SS14, km 163.5, Basovizza, Italy
12 Dipartimento di Fisica, Università di Pavia, via A. Bassi 6, Pavia, Italy
13 Sezione INFN di Pavia, Via A. Bassi 6, Pavia, Italy
14 Dipartimento di Matematica e Fisica, Università di Roma Tre, Via della Vasca Navale 84, Roma, Italy
15 Sezione INFN di Napoli e Dipartimento di Matematica e Fisica, Università della Campania "Luigi Vanvitelli", Viale Lincoln 5, Caserta, Italy
16 INAF-OAS Bologna, via P. Gobetti 93/3, Bologna, Italy
17 The Abdus Salam International Centre for Theoretical Physics, Strada Costiera 11, Trieste, Italy
18 Department de physique, Universitè de Lomè, Lomè, Togo
19 ISIS Neutron and Muon Source, STFC Rutherford-Appleton Laboratory, Didcot, OX11 0QX, United Kingdom
20 Riken Nishina Center, RIKEN, 2-1 Hirosawa, Wako, Saitama 351-0198, Japan
21 INO-CNR, via Madonna del Piano 10, 50019 Sesto Fiorentino, Italy
22 Università dell'Insubria, Dipartimento di Scienza e Alta Tecnologia, Via Valleggio 11, Como, Italy
23 Sezione INFN di Milano, via Celoria 16, Milano, Italy
24 Dipartimento di Fisica, Università degli Studi di Milano, via Celoria 16, Milano, Italy
25 IFN-CNR, Dipartimento di Fisica, Politecnico di Milano, piazza Leonardo da Vinci 32, Milano, Italy
26 Indian Centre for Space Physics, Kolkata, India
27 Dalian Institute of Chemical Physics of the Chinese Academy of Sciences, 457 Zhongshan Road, Dalian 116023, P. R. China

**Introduction**

The exploration of atomic structure has greatly advanced the progression towards the vision of modern physics. Thus, it is not unexpected that atoms are still being thoroughly studied, both theoretically and experimentally. Early time atomic spectroscopy allowed to unveil the relations describing electrons in the atomic system but also, to assess the nuclear structure and gave rise to the progressive establishing of the quantum field theory describing the interrelation between nucleus and the bound particles, Quantum Electrodynamics (QED) the most successful physics theory. In the study of the simplest atom, hydrogen, the description of the nucleus, the proton, and its interaction with the orbiting electron are at stake. A flow of theory hypotheses and experimental verification would occur if measurements and theory predictions conflict, leading unavoidably to a novel and deeper understanding of the system.

In this context, muonic atoms serve as the ideal testing ground to examine the limits between the theoretical foundations and the experimental results. Experiments on muonic atoms have enabled significant advances in the study of atomic and nuclear physics, in the tests of QED and of the Standard Model (SM). The muon, the heavier twin of the electron plays a significant role in the SM of particle physics and raises a persistent focus. Unlike the electron, which is stable, the muon decays through the weak force, the muon lifetime of 2.2μs permit us to make precision measurements to study atomic and nuclear properties as well as magnetic properties of condensed matter systems. The muon mass of about 106 MeV allows the muon to decay only into the electron and two neutrinos. High precision muonic atom experiments will further explore the accuracy frontier in particle physics, enabling the generation of significant data that will supplement those from high energy colliders. Muonic atoms are formed when negative muons, produced by particle accelerators with a kinetic energy of a few tens of MeV, are stopped in matter and captured by the Coulomb field of the nuclei; initially in excited states which promptly de-excite giving raise to the emission of characteristic X-rays. In muonic atoms the Bohr radius $r_n$ is inversely proportional to the orbiting particle's mass, the muon forms atomic states in which distances are reduced proportionally to the mass ratio between electron and muon. Being about 207 times smaller than normal atoms, muonic atoms give access to transitions which are rich of physical information and favorable to precision measurements with advanced experimental techniques Fig. 1. Furthermore, for high Z muonic atoms, the muon is inside of the atomic electron cloud for lowest quantum numbers and the Bohr radius for the 1s atomic state is well inside the nucleus. The changed Coulomb potential inside the nucleus causes the X-ray energy of the transitions to vary, providing information on the nuclear charge radius and allowing for accurate measurements of QED predictions.

The techniques that have been used throughout the time to measure muon-atomic transition lines with high precision are based either on crystal spectrometers, for atoms like Mg, Si, Li, C or on laser spectroscopy for light elements like H, D or He. There is an unavoidable correlation between the progress in theoretical calculations and the push towards more and more precise experiment where the theory, in general, leads the game. Precise prediction become the motivation for the experimental progression. In the 1980's, crystal spectrometer observations on muonic atoms confirmed and validated the most advanced theory predictions on the atomic transitions[1,2,3]. This led to a loss of incentive for the next incremental experimental endeavor towards better precision. The situation was all of a sudden greatly revived by the results coming from the experiment on Lamb shift in muonic hydrogen[4] (**pμ**$^-$) which allowed to derive the proton charge radius with unprecedented precision, contradicting measurements done with different techniques[5]. The energy levels of muonic hydrogen are orders of magnitude more responsive to the details of the proton structure than the levels of normal hydrogen. This long-chased laser spectroscopy measurement put in foreground the sensitivity of the muonic hydrogen system taken as an observatory able to serve, depending on the transition considered and the precision attained, as a cross check of predictions in tangent fields like QED, nuclear and particles physics. This experimental result gave rise to the iconic "proton radius puzzle" and sparked a decade-long flurry of theoretical and experimental analyses, all of which agreed that additional research into the muonic hydrogen atom system was necessary.

This occurrence immediately revived the previous idea to use laser spectroscopy to measure the *hyperfine splitting (hfs) in the ground state of muonic hydrogen*[6,7], see Fig. 1. The hyperfine splitting in muonic hydrogen represents a case where the accuracy of QED calculations exceeds the accuracy of the known values of fundamental physical parameters. Hence the measurement of $\Delta E^{1s}_{hfs}$ provide a unique possibility for the measurement of low energy proton magnetic structure with higher accuracy than what can be achieved in nuclear or particle physics experiments.

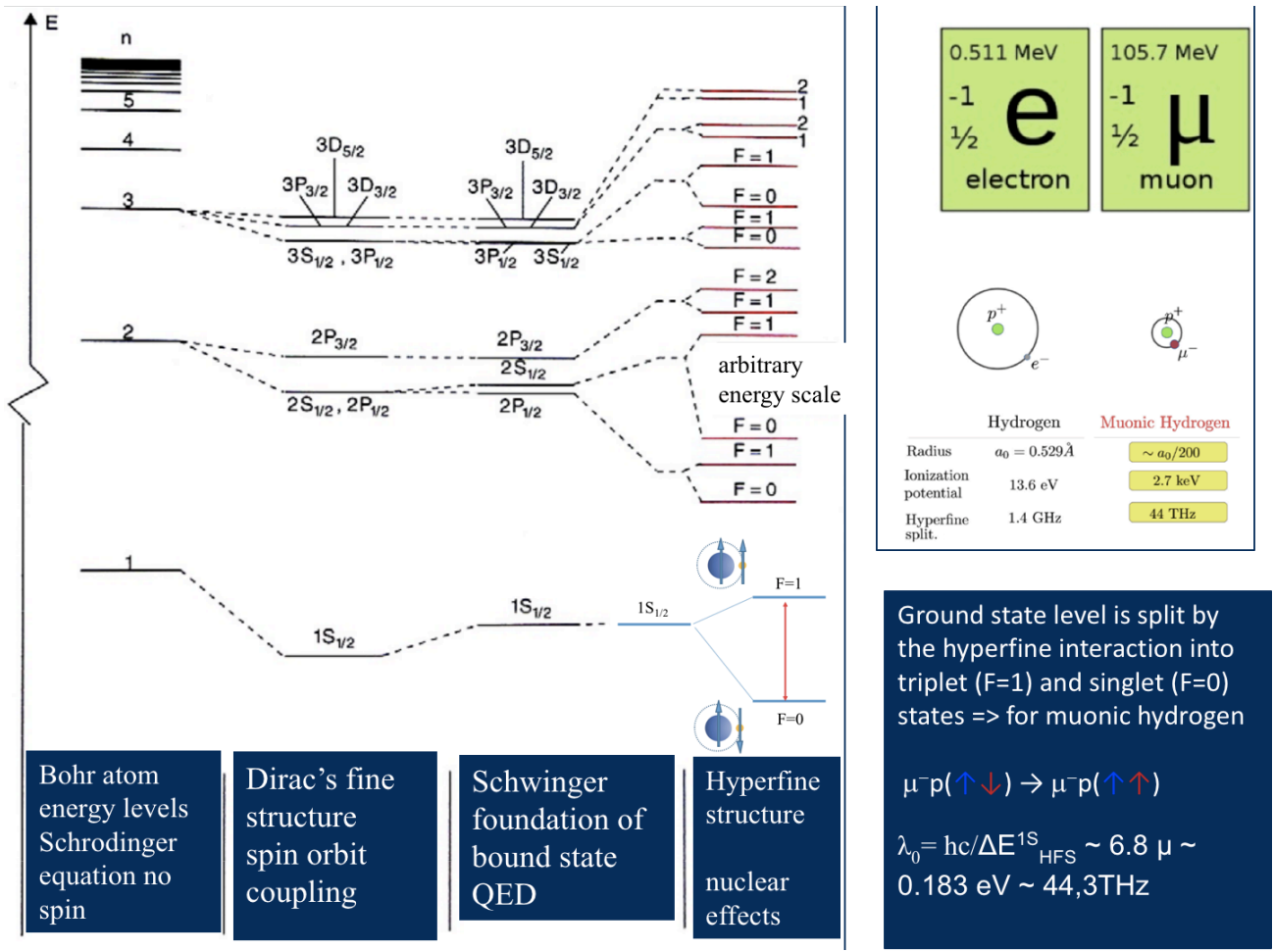

Fig. 1.
Scheme unfolding of the atomic structure evidencing the 1S hyperfine energy shift (right panel). Being approximately 200 times heavier than electron, the atomic radius of muonic hydrogen is about 1/200 of hydrogen. As a result, the negative muon in muonic hydrogen stays much closer to the proton than electron in ordinary hydrogen, and accordingly, the hyperfine splitting is much larger, the ground state hyperfine splitting of muonic hydrogen is 44 THz while in hydrogen is 1.4 GHz.

**Hyperfine splitting, why?**

The Zemach radius $R_p$ is the physical quantity related to the electromagnetic properties of the proton that can be extracted from the hfs measurements[7] it can be expresses as:

$$R_p = \left(\Delta E_{exp}^{hfs}/E^F - 1 - \delta^{QED} - \delta^{recoil} - \delta^{pol} - \delta^{hvp}\right)/(2m_{ep}\alpha),$$

where $E^F$ is the Fermi term expressed in terms of lepton and proton masses $m_l$, $m_p$ and the dipole magnetic moment of the proton $\mu_p$:

$$E^F = \frac{8}{3}\alpha^4 c^2 \frac{m_l^2 m_p^2}{(m_l + m_p)^3}\mu_p,$$

while $\delta^{QED}$, $\delta^{recoil}$, $\delta^{pol}$ and $\delta^{hvp}$ are correction terms related to the proton electromagnetic structure and to the strong interaction.

The Zemach radius of the proton is defined in terms of the first moment of the convolution of the charge $\rho_E(r)$ and magnetic moment distributions $\rho_M(r)$:

$$R_p = \int d^3 r\, r \int d^3 r'\, \rho_E(\boldsymbol{r} - \boldsymbol{r}')\rho_M(\boldsymbol{r}')$$

The µp's hyperfine splitting is the quantity that is most sensitive to the proton's Zemach radius. The experimental value of Rp places significant constraints on the parametrization of proton form factors as well as the theoretical models of proton electromagnetic structure[8][9], the measurement of the 1s hfs in the µp system is essential[10][11]. A method was originally draw up[6] at a time when the laser system couldn't be produced with the necessary qualities. A possible approach was subsequently identified by the FAMU collaboration[12][13].

**The experimental method**

Different experimental proposals for the measurement of ΔE$_{hfs}$ in the 1S state of the µp system have been put forth in recent years to provide top-accuracy data on the Zemach radius of the proton. This has been motivated by the need for new data on the proton electromagnetic structure that, as we have seen, had become a problem with the proton charge radius determination from the Lamb shift in muonic hydrogen.

The laser excitation of the ortho $F = 1$ hyperfine sub-level of the ground state of the muonic hydrogen atom from the para $F = 0$ sub-level (where F = F$p$ + Fµ is the total spin of the muonic hydrogen atom µ¯ $p$), is the challenging task that the three collaborations are tackling with various methodological approaches [14][15][13]. This is a very weak M1 magnetic dipole transition with probability P of only

P=2 × $10^{-5}$(E/J)(S/m$^2$)$^{-1}$(T/°K)$^{-1/2}$

with **E** laser pulse energy, **S** laser beam cross section and **T** target temperature [12]. In order to increase the likelihood of this laser-stimulated transition to a reasonable level, an optical multi-pass cavity must be used. In all these proposals the muonic hydrogen atom is being excited from the ground singlet to the triplet state with a laser, tunable around the resonance frequency ΔEhfs/h ∼ 44 THz; the experimental methods differ by the signature used to detect the laser-induced transitions Fig.2.

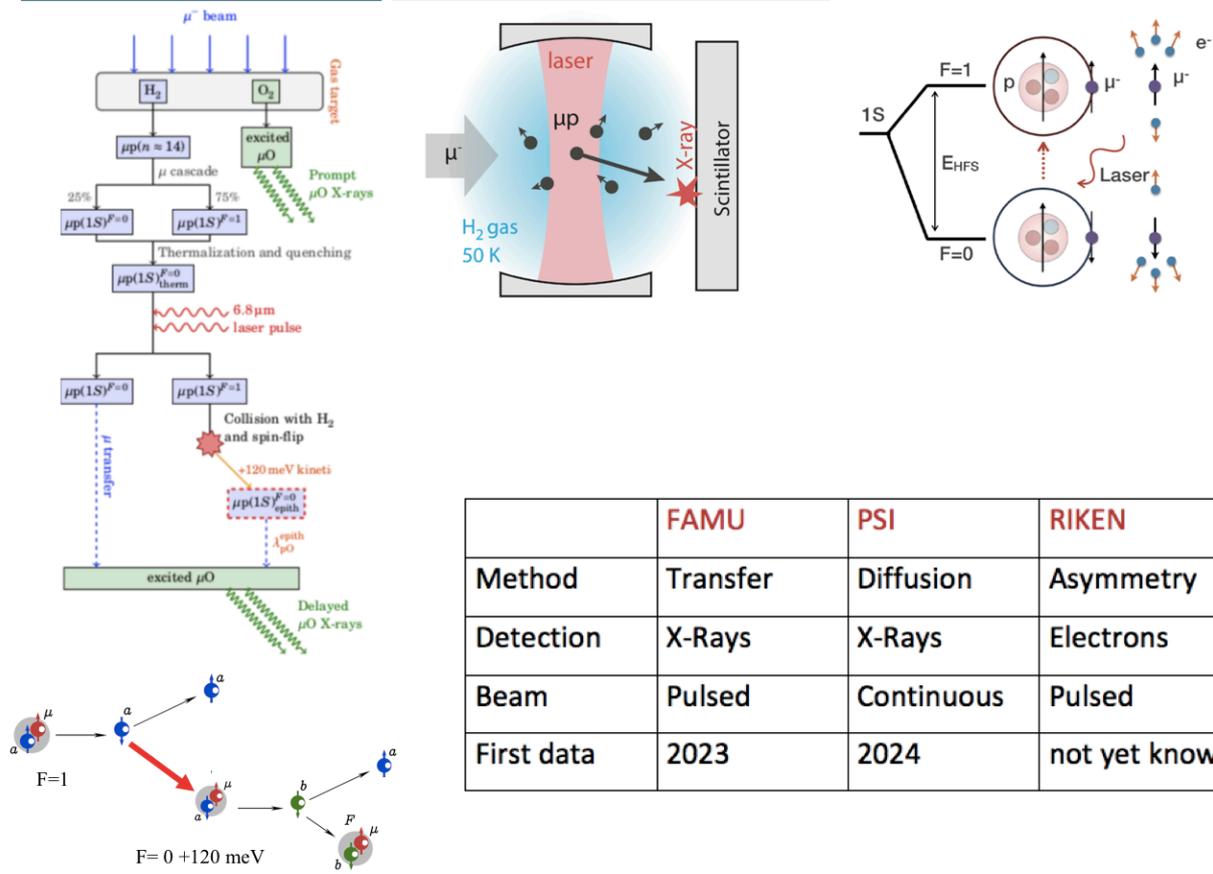

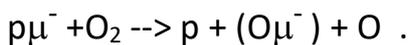

Fig. 2.
The three experimental approaches to the ΔEhfs measurement are represented here, left column the FAMU experiment at RAL (UK), central column the CREMA collaboration at PSI(CH), right column the RIKEN collaboration at KEK(J). The table summarizes the key aspects of each method.

The FAMU experimental method requires a detailed analysis of the succession of events that take place when muonic hydrogen atoms in a mixture of oxygen and hydrogen interact with laser radiation tuned to a frequency around the hyperfine transition resonance. Collisions of μp with oxygen lead to the reaction:

pμ⁻ + $O_2$ --> p + (Oμ⁻) + O .

The observable quantity used as signature is the *time distribution of the muon transfer events from hydrogen to oxygen* [16], which are signaled by the characteristic X-rays emitted during the de-excitation of the muonic oxygen The maximum deviation of the this X-ray time distribution from the time distribution in absence of the laser indicates that the laser source is tuned at the resonance frequency.
The μp atoms that have been excited to the triplet state with the laser pulse are accelerated, after the de-excitation in subsequent collisions with the surrounding $H_2$ molecules, by nearly 0.12 eV; the atoms carry the released energy away as kinetic energy. Since the rate of muon transfer to oxygen λpO(E) varies with the μp kinetic energy E, the observed time distribution of the characteristic X-rays is perturbed as compared to the time distribution in absence of laser radiation; the resonance frequency is recognized by the maximal response of the X-ray time distribution, Fig. 3. The efficiency of this method of detecting the events of laser-induced hyperfine excitation of the μp depends on how much the rate of muon transfer from accelerated μp atoms exceeds the transfer rate from thermalized atoms. The hydrogen-oxygen mixture had been selected for the FAMU method because of the evidence for a sharp energy dependence of λpO(E) at thermal and near epithermal energies (nearly an order of magnitude), that is not observed in other gases[17], Fig. 3.

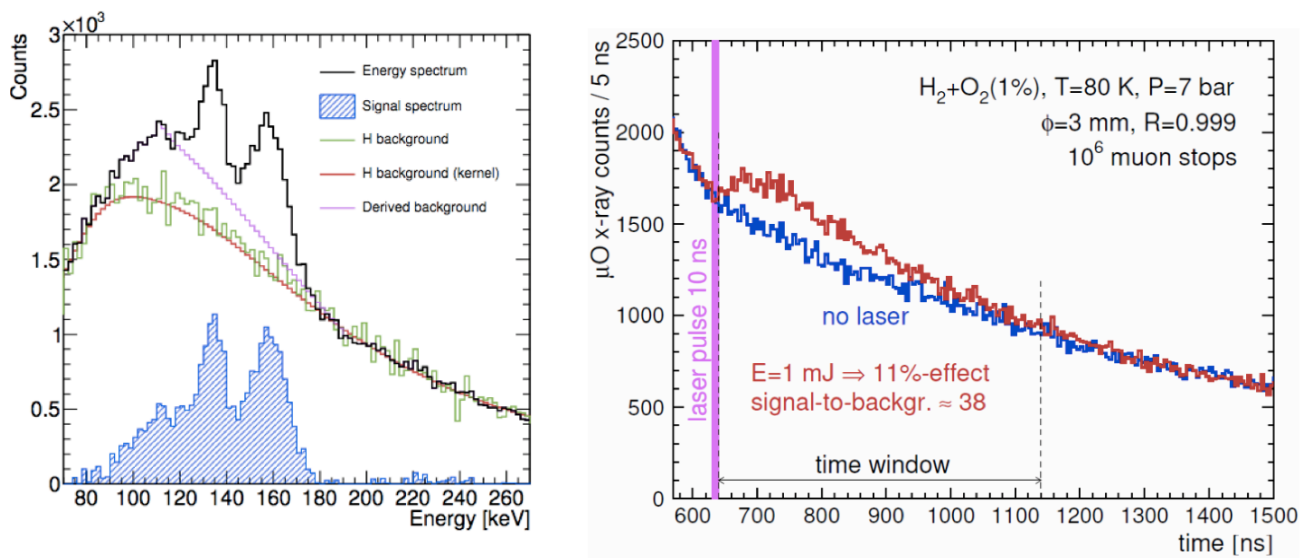

Fig. 3.
 Left panel: observed energy spectra of muonic oxygen K transition lines obtained in a mixture of hydrogen and oxygen gases (black line); pure hydrogen spectra are shown in green. Simulated time of arrival spectra of the delayed oxygen K lines with laser in ideal conditions at the appropriate hfs transition wavelength, and without laser are shown in the right panel, φ is the laser beam radius, R the cavity mirrors reflectivity.

The CREMA (PSI) collaboration employs a similar strategy, utilizing the gain in energy of the pμ atoms after deexcitation by drastically lowering the pure hydrogen gas target's size and pressure to enable the pμ atoms to reach a gold absorber to monitor the muonic gold X-rays temporal distribution.
The RIKEN collaboration has created a novel technique where the pure hydrogen gas target is kept at a low enough pressure for the muon decay to occur while the pμ is still in the F=1 spin state. Since for this experiment the laser light pumping the hyperfine transition is polarized, the spatial distribution of the muon decay electrons will make manifest the arisen F=0 -> F=1 transition.

## The FAMU laser solution

The pulsed muon source of the RIKEN-RAL muon facility [18], a tunable pulsed mid-infrared laser source tunable around the resonance energy of the 1S hfs of muonic hydrogen atom, $\Delta E_{hfs} \sim 0.182$ eV (6.78 µm), and a multi-pass cavity with extremely high reflectance mirrors will be crucial aspects of the FAMU experiment. In this mid-infra-red optical domain, no off-the-shelf laser system solution delivering the required light characteristics is available. A pulsed, tunable, narrow linewidth mid-infrared (mid-IR) laser radiation source was developed to meet the needs of the FAMU collaboration project. The laser system designed to fulfill those requirements is based on Difference Frequency Generation (DFG) using non-linear (NL) crystals pumped by two high-quality lasers with the first at 1064 nm and a second tunable around 1262nm. A photon is produced through DFG in the NL crystal from the difference in frequencies of two incident photon beams the pump and signal, Fig.4.

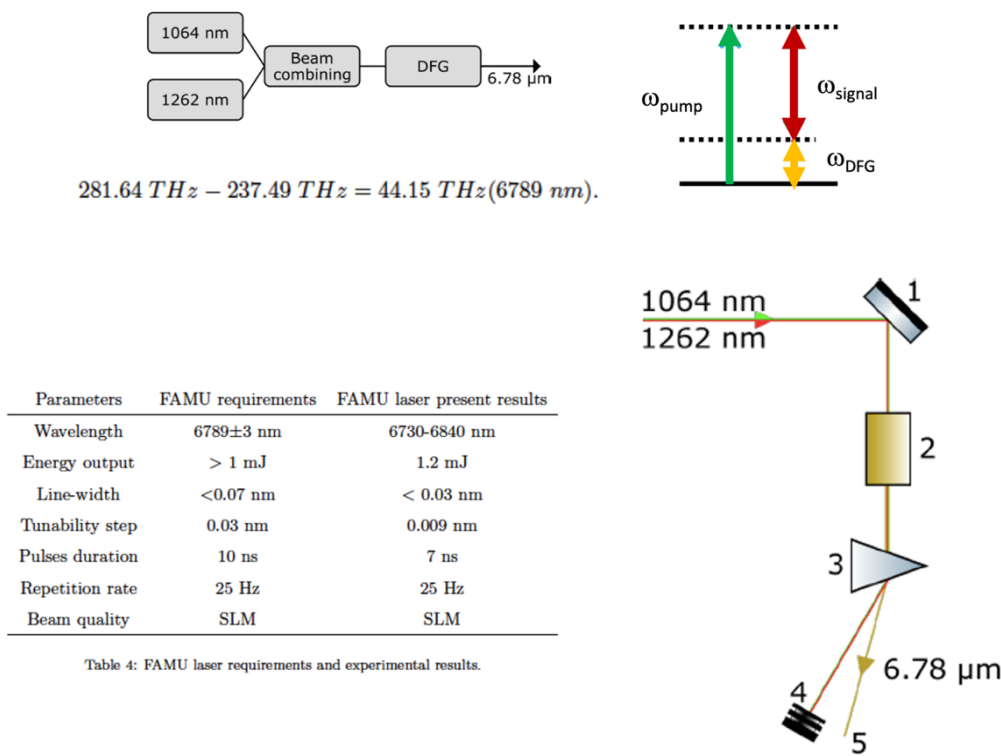

$$281.64\ THz - 237.49\ THz = 44.15\ THz\ (6789\ nm).$$

| Parameters | FAMU requirements | FAMU laser present results |
|---|---|---|
| Wavelength | 6789±3 nm | 6730-6840 nm |
| Energy output | > 1 mJ | 1.2 mJ |
| Line-width | <0.07 nm | < 0.03 nm |
| Tunability step | 0.03 nm | 0.009 nm |
| Pulses duration | 10 ns | 7 ns |
| Repetition rate | 25 Hz | 25 Hz |
| Beam quality | SLM | SLM |

Table 4: FAMU laser requirements and experimental results.

Fig. 4.
The FAMU laser system[19][20]; top panel the DFG explained in a block scheme, bottom the present laser characteristics; left panel diagram of DFG setup. In the figures are shown: 1 trichroic mirror, 2 NL crystal, 3 prism, 4 beam dump, 5 output light 6.78 µm. The system is continually evolving with the goal of increasing energy output while simultaneously keeping up with the progress of the manufacturing of DFG crystals.

## FAMU's experiment design

High precision spectroscopy is required to accurately identify this extremely weak transition signal and, in order to recover the signal, all potential background sources must be reduced to a minimum. This calls for thorough simulations before building the final layout. The pulsed beam of low energy muons will penetrate the cryogenic target full of high purity hydrogen with low oxygen contamination, primarily forming muonic hydrogen that quickly de-energizes and reaches the atomic ground level like what happens to all other muonic atoms generated by the stray muons.

A mosaic of specialized detectors covering the majority of the solid angle surrounding the target progressively records the flow X-rays, Fig. 5. The injection in the target of the laser radiation starts the actual measurement once the μp atoms have reached a thermal equilibrium with the surrounding gas. The multipass optical cavity that encloses the majority of the target's gas volume multiplies the laser beam's effects, Fig. 5. When the appropriate wavelength inducing the transition is attained, the time distribution of the muonic oxygen X-rays will abruptly change given the energy dependence of muon transfer from μp to oxygen.

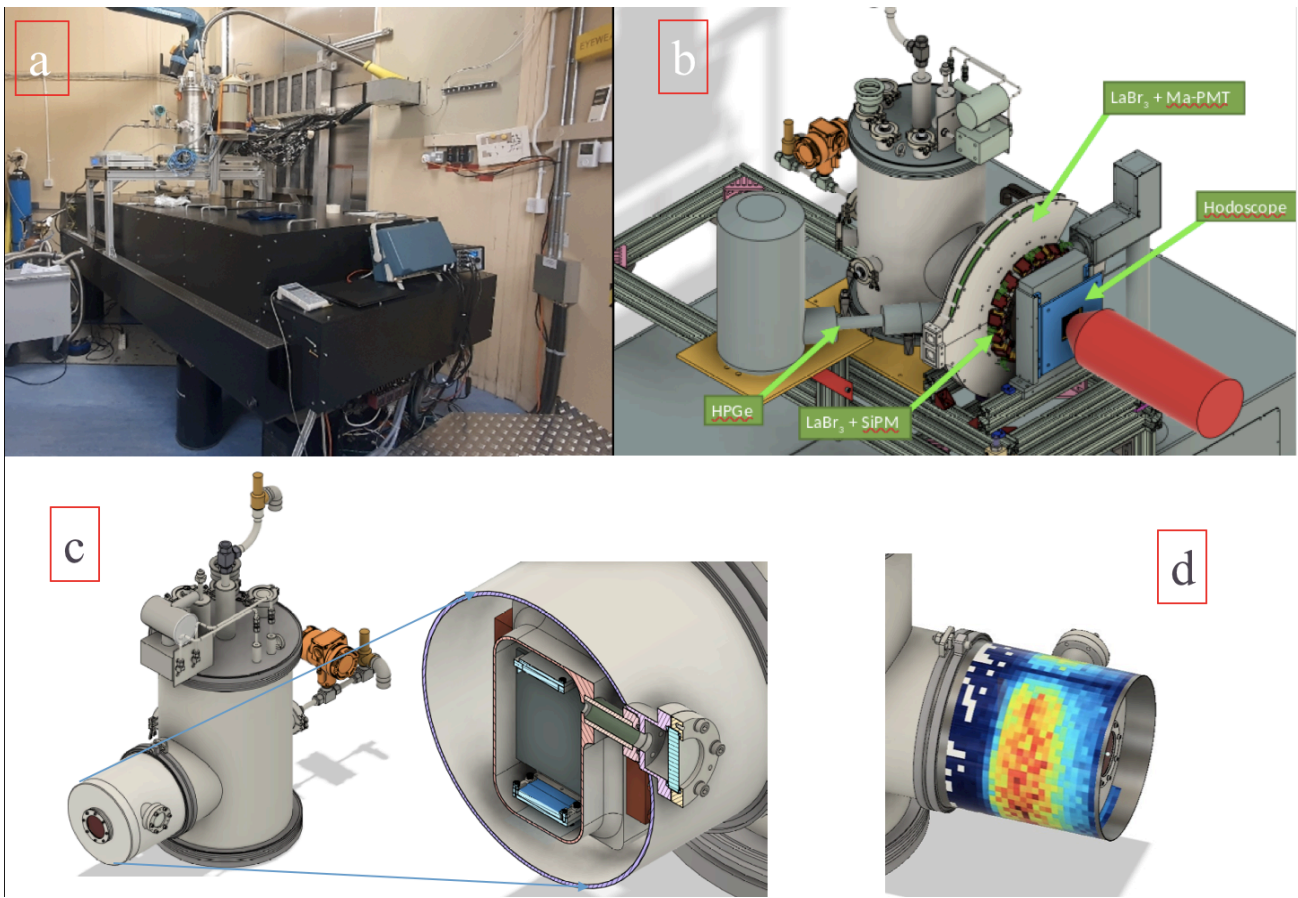

Fig. 5.
Visible in a) a picture of the FAMU experiment lay-out at the Port 1 of the RIKEN RAL facility, the whole experiment is placed on a large optical table which supports enclosed in a box the laser and cryo-target wit the detection system. The beam exits the gray led wall in the upper central part of the picture. In b) a CAD design of the target and the detection system. The c) panel shows a detail of the cryo-targed and an expanded section evidencing the multipass cavity. In c) a simulatin of the oxygen X-rays impact points corresponding with the location of the detection system.

The dedicated and harmonious work of the FAMU international team's individual members is vital. In Fig. 6, we see the majority of the Italian team members at a meeting at Milano-Bicocca University, other team members participating remotely.

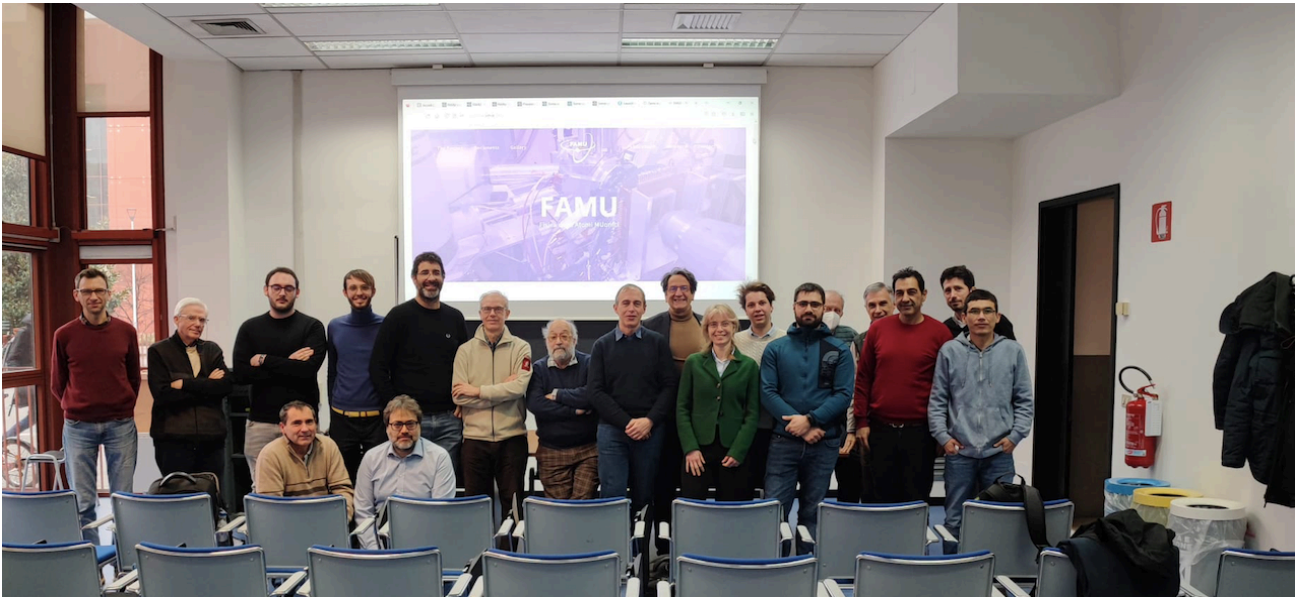

Fig. 6.  Members of the FAMU team.